\documentclass[aps,prb,twocolumn,showpacs]{revtex4-1}
\usepackage{amsfonts}
\usepackage{amssymb}
\usepackage{amsmath}
\usepackage{bm}
\usepackage[dvips]{graphicx}
\usepackage[dvips]{color}
\usepackage{graphicx}
\usepackage{epsfig}
\newcommand{\beq}{\begin{equation}}
\newcommand{\eeq}{\end{equation}}
\newcommand{\beqa}{\begin{eqnarray}}
\newcommand{\eeqa}{\end{eqnarray}}

\begin{document}
\title{Spin Relaxation due to Charge Noise}
\author{Peihao Huang}
\author{Xuedong Hu}
\email{xhu@buffalo.edu}
\affiliation{Department of Physics, University at Buffalo, SUNY, Buffalo, NY 14260, USA}

\begin{abstract}
We study decoherence of an electron spin qubit in a quantum dot due to charge noise. We find that at the lowest order, the pure dephasing channel is suppressed for both $1/f$ charge noise and Johnson noise, so that charge noise leads to a pure relaxation channel of decoherence. Because of the weaker magnetic field dependence, the spin relaxation rate due to charge noise could dominate over phonon noise at low magnetic fields in a gate-defined GaAs or Si quantum dot or a InAs self-assembled quantum dot. Furthermore, in a large InAs self-assembled quantum dot, the spin relaxation rate due to phonon noise could be suppressed in high magnetic field, and the spin relaxation due to charge noise could dominate in both low and high magnetic field. Numerically, in a 1 Tesla field, the spin relaxation time due to typical charge noise is about $100$ s in Si, $0.1$ s in GaAs for a gate-defined quantum dot with a $1$ meV confinement, and $10$ $\mu$s in InAs self-assembled quantum dot with a $4$ meV confinement.

%The relaxation rate depends linearly on the applied magnetic field for $1/f$ noise, and is inversely proportional to the fourth power of the dot confinement energy.
\end{abstract}

\date{\today}
\maketitle

\section{Introduction}

The experimental and theoretical investigation of spin qubits have seen impressive progress in recent years.\cite{Hanson2007,Morton2011,Zwanenburg2013} Experimentally, initialization, manipulation and detection have all been demonstrated for single electron spin qubit in quantum dots \cite{Hanson2007,Morton2011,Zwanenburg2013} and donors.\cite{Morello2010,Pla2012} Partial to full electrical control have also been demonstrated for logical qubits encoded in two- or three-spin states.\cite{Petta2005, Bluhm2011, Maune2012, Shulman2012, Gaudreau2012,Medford2013_preprint}

Decoherence is one of the key indicators of whether a physical system can act as a qubit. Decoherence of a single electron spin in a finite field is mainly due to the hyperfine (HF) interaction induced pure dephasing,\cite{Khaetskii2002, Witzel2006, Yao2006, Deng2006, Cywinski2009, Barnes2012} although this pure dephasing channel can be alleviated by spin-echo and more sophisticated dynamical decoupling techniques,\cite{Petta2005,Bluhm2011,Maune2012} or nuclear bath polarization and purification.\cite{Ramon2007,Stich2007,Huang2010} Ultimately, the limit to spin coherence is set by spin relaxation. Two main spin relaxation channels have been studied so far, one due to electron-phonon interaction and spin-orbit (SO) interaction,\cite{Khaetskii2001, Golovach2004, Tahan2005} the other due to electron-phonon interaction and hyperfine interaction.\cite{Erlingsson2002} The first one is generally the strongest relaxation channel, with the relaxation rate having a $B_0^{5}$ (or $B_0^{7}$) dependence with the applied magnetic field when the piezoelectric (or deformation) phonon noise dominates.\cite{Golovach2004}

Charge noise is ubiquitous in nanostructures including semiconductor and superconductor devices.\cite{Jung2004,Muller2006,Buizert2008,Hitachi2013,Takeda2013} It poses a significant challenge to the charge sensitive qubit schemes, such as charge qubits \cite{Petersson2010, Dovzhenko2011, Astafiev2004, Schreier2008, Pashkin2009} and $S-T_0$ qubit.\cite{Coish2005, Hu2006, Ramon2012, Dial2013} Charge noise in a semiconductor heterostructure device could come from a variety of sources, for example, the $1/f$ noise from dynamical traps (most probably near the various interfaces), Johnson noise and evanescent wave Johnson noise (EWJN) from the metallic gates, etc.\cite{Galperin2006, Marquardt2005, SanJose2006, SanJose2007, Langsjoen2012, Poudel2013} In the case of a single electron spin qubit, although the electron spin does not directly couple to the charge fluctuations, SO interaction does allow charge noise to induce spin decoherence. Existing works show that Johnson noise from the metallic gates could be important for single electron spin relaxation in a GaAs gate-defined quantum dot (QD) when the magnetic field is weak.\cite{Marquardt2005, SanJose2006, SanJose2007} In addition, through the direct magnetic dipole interaction, EWJN could be a important spin relaxation channel when surface metallic gates are sufficiently close to the confined electron.\cite{Langsjoen2012,Poudel2013}

In this paper, we present a comprehensive study of spin decoherence of a quantum-dot-confined electron due to charge noise through SO interaction, where the QD includes GaAs and Si gated-defined QD and InAs self-assembled QD (SAQD), and the charge noise includes the $1/f$ noise, Johnson noise and EWJN. Since charge noise, such as $1/f$ noise, is most important at low frequencies, we modify the existing studies of SO interaction by accounting for the field induced QD displacement, and the modified treatment could be easily extended to more complex situations, such as an electron in a moving QD. \cite{Hermelin2011, McNeil2011, Huang2013} We find that charge noise can induce both relaxation and pure dephasing, although the latter turns out to be very weak, so that practically charge noise leads to a pure relaxation channel for spin decoherence in most situations. Furthermore, the spin relaxation rate due to charge noise could dominate over phonon noise at low magnetic fields in a gate-defined GaAs or Si quantum dot or a InAs SAQD; In a large InAs SAQD, the spin relaxation rate due to phonon noise could be suppressed in high magnetic field, and the spin relaxation due to charge noise could dominate in both low and high magnetic field.

\section{Theoretical Formalism}
\subsection{System Hamiltonian}

\begin{figure}[tb]
\centering
\includegraphics[scale=1.2]{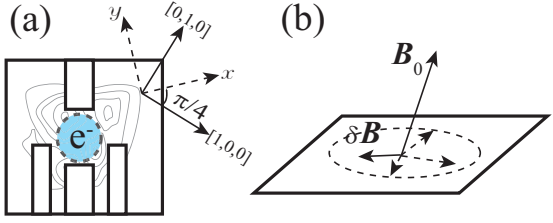}
\caption{A schematic of a spin qubit in a gate confined QD. Panel (a) gives the topview of the structure and the coordinate system ($xyz$) defined in the laboratory frame, with $x$ and $y$ along the [$110$] and [$\bar{1}10$] directions. Panel (b) gives the sideview and the effective magnetic field.}\label{fig1}
\end{figure}

The system we consider is a single electron in a gate-defined QD, as shown in Fig.~\ref{fig1}. In general, the growth-direction ([001]-direction in this paper) confinement is much stronger, so that the vertical momentum fluctuation is strongly suppressed compared to the in-plane fluctuations. Therefore, we focus on the electron dynamics in the in-plane directions, with the QD modeled as a 2D harmonic potential. The Hamiltonian for the QD-confined electron in the presence of SO interaction and charge noise is
\begin{eqnarray}
H &=&H_{d}+H_{Z}+H_{SO},  \label{H} \\
H_{d} &=&\frac{{\pi }^{2}}{2m^{\ast }}+V\left( \boldsymbol{r}\right) +\delta V\left( \boldsymbol{r},t\right) ,  \label{Hd} \\
H_{Z} &=&\frac{1}{2}g\mu _{B}{\boldsymbol{B}}_{0}\cdot {\boldsymbol{\sigma }},  \label{HZ} \\
H_{SO} &=&\beta _{-}\pi _{y}\sigma _{x}+\beta _{+}\pi _{x}\sigma _{y}.
\label{Hso}
\end{eqnarray}
The subscripts $d$, $Z$, and $SO$ refer to "dot", "Zeeman", and "spin-orbit". In $H_{d}$, ${\boldsymbol{\pi }}$ is the electron 2D momentum ($e>0$), given by $\boldsymbol{\pi }=-i\hbar {\boldsymbol{\nabla }}+(e/c)\boldsymbol{A}(\boldsymbol{r})$, and $V\left( \boldsymbol{r}\right) $ is the static confinement potential of the QD, which is assumed to be harmonic $V\left( \boldsymbol{r}\right) =\frac{1}{2}m^{\ast }\omega _{d}^{2}r^{2}$; $\delta V\left( \boldsymbol{r},t\right) $ captures the charge noise in the system, which is $\delta V\left( \boldsymbol{r},t\right) =\delta V\left(0,t\right) -e\boldsymbol{E}_{c}(t) \cdot \boldsymbol{r}$, where $\boldsymbol{E}_{c}(t) =-\boldsymbol{\ \nabla }\delta V\left( 0,t\right) /e$ ($e>0$) is the electric field of the charge noise. In $H_{Z}$, ${\boldsymbol{B}}_{0}$ is the applied magnetic field (with $\boldsymbol{\hat{n}}_{0}$ its unit vector). In $H_{SO}$, $\beta _{\pm }\equiv \left( \beta \pm \alpha \right) $, where $\alpha $ and $\beta $ are the Rashba and Dresselhaus SO interaction constants. The $x$ and $y$ axes are along the [$110$] and [$\bar{1}10$] directions. If $x$ and $y$ had been defined along the [$100$] and [$010$] directions, the SO term would have taken the usual form $H_{SO}=\beta (-\pi_{x}\sigma _{x}+\pi _{y}\sigma _{y})+\alpha (\pi _{x}\sigma _{y}-\pi_{y}\sigma _{x})$.\cite{Golovach2004,Borhani2006,Huang2013} The current choice of $x$ and $y$ helps simplify the presentation below.

\subsection{Effective Spin Hamiltonian}

The key to the study of decoherence of an electron spin is to disentangle the spin dynamics from the orbital dynamics. To achieve this separation, we perform a Schrieffer-Wolff transformation $\tilde{H}=\exp (S)H\exp (-S)$, and require $\left[ H_{d}+H_{Z},S\right] =H_{SO}$ to remove the SO Hamiltonian in the leading order.\cite{Golovach2004, Borhani2006,Aleiner2001, Stano2006, Huang2013}  Here the time-dependent potential $\delta V\left( r,t\right) $ is included in $H_d$, rather than treated as a perturbation, as compared with previous studies.\cite{Golovach2004,Borhani2006,Huang2013} The reason for this change is that charge noise is more important at low frequencies (and the electron Zeeman energy is much smaller than the orbital excitation energy) and long wave length (so that the corresponding electric field is uniform in a QD).  Motion of the QD due to charge noise is thus adiabatic, and the QD harmonic potential is centered at a position determined by the instantaneous total potential (from the gates and the charge noise).  The physical picture is quite clear here: charge noise causes the QD potential (and therefore the electron, which stays in the ground orbital state of the QD) to wander around its designated position, and through the SO interaction the spin would sense this wandering in the form of a magnetic noise.

Defining superoperator $\mathbb{L}_{x}$ as $\mathbb{L}_{x}A\equiv \lbrack H_{x},A]$ (with $x=d$ or $Z$), the condition $\left[ H_{d}+H_{Z}, S \right] = H_{SO}$ on $S$ can be formally expressed as\cite{Golovach2004, Borhani2006, Huang2013}
\begin{equation}
S=\sum_{m=0}\left( \frac{\mathbb{L}_{Z}}{\mathbb{L}_{d}}\right) ^{m}\mathbb{L}_{d}^{-1}H_{SO} .
\end{equation}
%where Zeeman energy is usually much less than the dot confinement energy: $\omega_Z \ll \omega_d$.
With a harmonic confinement $V\left(r\right) =\frac{1}{2}m^{\ast }\omega _{d}^{2}r^{2}$ from the external gates, the total instantaneous QD potential, including the electric field from the charge noise, is $\frac{1}{2}m^{\ast }\omega _{d}^{2}\left[\boldsymbol{r}-\boldsymbol{R}\left( t\right) \right] ^{2}+ {\rm const}$, where $\boldsymbol{R}(t) ={e\boldsymbol{E}_{c}(t) }/({m^{\ast }\omega _{d}^{2}})$ is the displacement due to charge noise. After some algebra, we obtain (see Appendix \ref{append::Ld})
\begin{equation}
\mathbb{L}_{d}^{-1}H_{SO} = i\left(\boldsymbol{\sigma }\cdot\boldsymbol{\xi }\right), \label{LdHso}
\end{equation}
where $\boldsymbol{\xi }$\ is a vector in the 2DEG plane,
\begin{equation}
{\boldsymbol{\xi }}\left( t\right) \equiv m^{\ast }/\hbar \left[ \beta_{-}\left( y-R_{y}\right) ,\beta _{+}\left( x-R_{x}\right) ,0\right]. \label{xi}
\end{equation}
%We retain $S$ only to the lowest order, assuming the energy scales satisfy
Due to the motion of the QD potential, the vector $\boldsymbol{\xi }(t)$ is time-dependent. Under most circumstances, conditions $m^{\ast }\left( \beta ^{2}+\alpha^{2}\right) \ll \hbar \omega _{Z}\ll \hbar \omega _{d}$ ($\omega _{Z}$ is the Zeeman frequency) are satisfied. For example, in GaAs the SO coupling energy $m^{\ast }\left( \beta ^{2}+\alpha^{2}\right)$ is in the order of $1 \mu$eV, the Zeeman splitting $\hbar\omega_Z$ is about 25 $\mu$eV per Tesla, and the orbital confinement energy $\hbar \omega_d$ is in the order of 1 meV.  Under these conditions, the Schrieffer-Wolff transformation matrix $S$ can be simplified to
\begin{equation}
S\left( t\right) \approx i\boldsymbol{\sigma }\cdot \boldsymbol{\xi }\left(t\right), \label{S}
\end{equation}
and the transformed Hamiltonian takes the form
\begin{equation}
H^{\prime }=i\hbar \partial _{t}S+H_{d}+H_{Z}+[S,H_{SO}]+\frac{1}{2!}[S,[S,H_d+H_Z]]+\cdots. \nonumber
\end{equation}
%The displacement of the QD potential $\boldsymbol{R}$ is the only time dependent terms in Eq. (\ref{S}),
%\begin{equation}
%\partial _{t}i\boldsymbol{\xi }=-i\frac{m^*}{\hbar}\left[ \beta_{-}\dot{R}_{y},\beta_{+}\dot{R}_{x},0\right].
%\end{equation}
%We consider the adiabatic condition, where the energy scale of electron motion is much less than dot confinement energy $E_d$, so that the electron orbital state stays in the instantaneous ground
%state $\psi (\boldsymbol{r})=\exp\left( -(\boldsymbol{r}-\boldsymbol{R})^2/2\lambda ^{2}\right) /\lambda \sqrt{\pi }$, %where $\lambda ^{-2}=\hbar ^{-1}\sqrt{(m^{\ast }\omega _{d})^{2}+(eB_{z}/2c)^{2}}$, up to a
%magnetic phase that does not affect the calculation of the %effective magnetic fields. Then,
At the 1st order of SO interaction, the last two terms in $H^\prime$ can be neglected as they are second order in $H_{SO}$.  Furthermore, $H_d$ is now decoupled from the spin dynamics, so that we obtain an effective spin Hamiltonian
\begin{eqnarray}
H_{eff} &=&\frac{1}{2}g\mu _{B} \left[ \boldsymbol{B}_{0}+\delta \boldsymbol{B}(t) \right]\cdot \boldsymbol{\sigma },  \label{Heff} \\
\delta \boldsymbol{B}(t) &=&\frac{2}{g\mu _{B}}\frac{e}{\omega _{d}^{2}}\left[ \beta _{-}\dot{E}_{cy}\left( t\right) ,\beta _{+}\dot{E}_{cx}\left( t\right) ,0\right] .  \label{dB1}
\end{eqnarray}
In this spin Hamiltonian, the time derivative of the charge-noise-induced random electric field, $\dot{E}_{c}\left( t\right)$, leads to an effective magnetic noise $\delta \boldsymbol{B}\left( t\right) $ for the electron spin. This conversion is through the dot motion $\boldsymbol{R}(t)$ and the SO interaction. Indeed, $\delta \boldsymbol{B}\left( t\right) $ coincides with the SO interaction term $H_{SO}$ if the momentum operator $\boldsymbol{\pi }$ in Eq.~(\ref{Hso}) is substituted by the drift momentum $m^{\ast }\partial _{t}\boldsymbol{R}\left( t\right)$, where $\partial_{t} \boldsymbol{R} \left( t\right)$ is the drift velocity of the QD-confined electron.  Equation~(\ref{dB1}) also shows that the magnetic noise in general has both longitudinal and transverse components (relative to the total field that acts as the quantization axis), which could induce both relaxation and pure dephasing for the electron spin qubit. This result is different from the previous treatment in the phonon case, where pure relaxation process is obtained.\cite{Golovach2004}  However, we would like to emphasize that generally the pure dephasing we obtain is much slower than relaxation, so that effectively charge noise leads to a pure relaxation channel for spin decoherence.

The approach we adopt here is at the same order of perturbation theory as the conventional approach,\cite{Golovach2004, Borhani2006, Stano2006, Huang2013} where phonon noise and SO interaction are both taken to the first order in the perturbative calculation.  In fact, we have adapted our approach to the case of phonon noise and obtained similar results as in previous works.  By including charge noise in $H_0$, we slightly shifted the overall quantization scheme, so that our magnetic noise does not have the neat transversal form derived in previous works.  As a reward, we obtain an appealingly simple physical picture, in which the magnetic noise arises from the wandering QD, while the electron stays in the ground orbital state.  If we adopt the conventional approach, our QD would have been fixed at the gate-designated spot, and the random electric field from the charge noise would cause the electron to be excited into the P-orbitals.  In effect, in the conventional approach we have a fixed QD and a wandering electron in the dot, while in our approach we have a wandering QD with a ``grounded'' electron. The present approach introduced here can be easily extended to a moving quantum dot case where the QD potential has a finite displacement away from the original minimum, which facilitate further exploration for more complex situations. \cite{Hermelin2011, McNeil2011, Huang2013}

\subsection{Noise Correlation}

To calculate the spin relaxation rates due to charge noise, we need to obtain the correlation functions $\left\langle \delta B_{i}\delta B_{j}\left( t\right) \right\rangle $ of the magnetic noise, or the correlation functions $\langle \dot{E}_{ci}\dot{E}_{cj}\left( t\right)\rangle $ of the time derivative of the random electric field. We assume charge noise is isotropic and have time-translational symmetry
\begin{equation}
\langle E_{ci}(t_{1})E_{cj}(t_{2})\rangle =\delta _{ij}S_{E}(t_{2}- t_{1}),
\end{equation}
where $\delta _{ij}$ is the Kronecker delta function ($i,j=x$ or $y$). Suppose $S_{E}(\omega)$ is the Fourier transform of $S_{E}(t)$, then
$\langle \dot{E}_{ci}(t_{1})\dot{E}_{cj}(t_{2})\rangle = \delta _{ij}\frac{1}{2\pi }\int_{-\infty }^{+\infty }d\omega S_{E}(\omega)\omega^{2} e^{-i\omega(t_{2}-t_{1})},
$ so that the Fourier transform of $\langle \dot{E}_{ci}(t_{1})\dot{E}_{cj}(t_{2})\rangle$ is
\begin{equation}
\langle \dot{E}_{ci}\dot{E}_{cj}\rangle_{\omega} = \delta _{ij} S_{E}(\omega)\omega^{2}.
\end{equation}
Thus, as long as the spectrum $S_{E}(\omega)$ for electric field is given, the corresponding spectrum $\langle \dot{E}_{ci}\dot{E}_{cj}\rangle_{\omega} $ is also known. Below we examine three types of charge noise in semiconductor nanostructures.

\textit{$1/f^a$ charge noise---}$1/f$ charge noise is ubiquitous in solid state materials, and semiconductor nanostructures are no exception. \cite{Dutta1981,Weissman1988} In general, the frequency dependence is not exactly $1/f$, but $1/f^a$,where the exponent $a$ ranges between 0 and 2.\cite{Dutta1981,Weissman1988,Muller2006,Buizert2008} In the following, we will use the name $1/f^a$ noise in general, and $1/f$ noise is reserved for the case $a=1$. Thus, the electric field correlation of $1/f^a$ charge noise is
\begin{equation}
S_{E}\left( \omega \right) = \frac{A}{\omega ^{a}}.
\end{equation}
The parameter $A$ can be related to the energy level fluctuation $\sigma_V$ of a QD. Normally, the measured energy fluctuation is dependent with the frequency range in those experiments.\cite{Jung2004, Buizert2008, Hitachi2013} If the noise spectrum is assumed to be 1/f, then a frequency independent quantity can be defined $\widetilde{\sigma}_V=\sigma_V/\sqrt{\ln(\omega_c/\omega_0)}$, where $\omega_0$ and $\omega_c$ are the low and high frequency limit. Then, the effective energy level fluctuations $\widetilde{\sigma}_V$ can be estimated, which ranges from $0.1$ $\mu$eV to 1 $\mu$eV based on the experimental measurements.\cite{Jung2004, Buizert2008, Hitachi2013, Petersson2010} This energy fluctuation is due to fluctuations in the electrical voltage at the QD. Assuming that the voltage fluctuation comes from the charging and discharging of traps near interfaces, we conservatively estimate the electric field strength as $\sigma _{E}=\sigma _{V}/\left( el_{0}\right) =1$ V/m, so that $A=\sigma_{E}^{2}=1$ (V/m)$^{2}$ with $a=1$. Here $\widetilde{\sigma}_V=0.1$ $\mu$eV is assumed, and the length scale $l_{0}$ between the QD and the traps is chosen as 100 nm, which is a typical barrier thickness for a QD, considering that $1/f$ noise most probably comes from traps near the interfaces between the metallic gates and the barrier material.

\textit{Johnson Noise---}Johnson Noise is always present in electrical circuits. Since our QD is gate-defined, Johnson noise also affects the electron spin. Its spectrum $S_{V}\left( \omega \right) =\int_{-\infty }^{+\infty }\left\langle \delta V\left( 0\right) \delta V\left( t\right) \right\rangle \cos \left( \omega t\right) dt$ is \cite{Weiss1999}
\begin{equation}
S_{V}\left( \omega \right) =\frac{2\xi \omega \hbar ^{2}}{1+\left( \omega/\omega _{R}\right) ^{2}}\coth \left( \hbar \omega /2k_{B}T\right), \label{Sv_John}
\end{equation}
where $\xi =R/R_{k}$ is dimensionless constant, $R_{k}=h/e^{2}=26$ k$\Omega $ is the resistance quantum, $R$ is the resistance of the circuit, and $\omega _{R}=1/RC$ is the cutoff frequency. Suppose the Johnson from the circuits outside the dilution refrigerator are strongly filtered, and we consider only the Johnson noise from the circuits inside the dilution refrigerator with the resistance being 50 Ohm. Based on the same argument as for $1/f^a$ charge noise, we have,
\begin{equation}
S_{E}(\omega) =S_{V}(\omega)/(el_{0})^{2},
\end{equation}
where, $l_{0}$ is the length scale chosen as 100 nm.

\textit{Evanescent Waves Johnson Noise---}When the QD is sufficiently close to the metal gates, the evanescent waves from the metallic gates give rise to additional electric fluctuations, which depends on the distance between the QD and the gates.
Based on the local electrodynamics and the quasistatic approximation, the spectrum of the electric field from EWJN is given by Ref.~\onlinecite{Langsjoen2012,Poudel2013}
\begin{equation}
S_{E}(\omega) = \frac{\hbar}{16\epsilon_0z^3} \mathrm{Im} \frac{\epsilon-1}{\epsilon+1}\coth \left( \hbar \omega /2k_{B}T\right),\label{EWJNSE}
\end{equation}
where $\epsilon_0$ is the vacuum permittivity, $z$ is the distance between the QD and the metal gates and $\epsilon$ is the relative permittivity. In the case of high conductivity of the gate, we have $\epsilon\approx i\frac{\sigma}{\omega\epsilon_0}$, so that\cite{Poudel2013}
\begin{equation}
S_{E}(\omega) = \frac{\hbar\omega }{8z^3\sigma}\coth \left( \hbar \omega /2k_{B}T\right),
\end{equation}
where, $\sigma$ is the conductivity of the metallic gate. Local electrodynamics is applicable when $z$ is large than the QD size and quasistatic approximation is valid when $z$ is less than tenth of the skin depth $\delta$ of the metal.\cite{Langsjoen2012,Poudel2013} For copper near absolute zero, $\delta\sim 3$ $\mu m$. In this paper, we choose $z=100$ nm, where the expression in Eq. (\ref{EWJNSE}) is valid.

\subsection{Spin Relaxation Rates}

The decoherence of the electron spin $\boldsymbol{S}={\boldsymbol{\sigma }}/2$ is governed by Hamiltonian (\ref{Heff}). In the regime where the noise correlation time is much shorter than the spin decay time, the dynamics and relaxation of the spin is governed by the Bloch equation.\cite{Slichter1980}
% (see Appendix \ref{append::blocheqn})
To simplify our calculation, we first rotate to a new $(XYZ)$ coordinate system, in which $Z$ axis is along the direction of the applied magnetic field.  The relaxation and dephasing time $T_{1}$ and $T_{2}$ are then given by \cite{Borhani2006, Slichter1980}
%(see also Appendix \ref{append::spindecoher})
\begin{eqnarray}
\frac{1}{T_{1}} &=&S_{XX}(\omega _{Z})+S_{YY}(\omega _{Z}), \\
\frac{1}{T_{2}} &=&\frac{1}{2T_{1}}+ \left. S_{ZZ}(\omega)\right|_{\omega\rightarrow0},
\end{eqnarray}
where, the correlation functions are
\begin{equation}
S_{ij}(\omega )=\frac{g^{2}\mu _{B}^{2}}{2\hbar ^{2}}\int_{-\infty }^{+\infty }\left\langle\delta B_{i}(0)\delta B_{j}(t)\right\rangle \cos(\omega t) dt.
\end{equation}

As a general example, let us consider a magnetic field $\boldsymbol{B}_{0}=B_{0}\left(\sin\theta\cos \varphi ,\sin\theta\sin \varphi ,\cos\theta\right)$ in an arbitrary direction, where $\theta $ and $\phi $ are the polar and azimuthal angles of the magnetic field in the ($xyz$) coordinate system. We rotate ($xyz$) to ($XYZ$) coordinate system, so that $Z$-axis is along the direction of $\boldsymbol{B}_{0}$. Correspondingly, the axis $\hat{X}$, $\hat{Y}$ and $\hat{Z}$ in the original $(xyz)$ coordinate frame is
\begin{eqnarray}
\hat{X}&=&\left(\cos\theta\cos\varphi,\cos\theta\sin\varphi,-\sin\theta\right)^T, \\
\hat{Y}&=&\left(-\sin\varphi,\cos\varphi,0\right)^T,\\
\hat{Z}&=&\left(\sin\theta\cos \varphi ,\sin\theta\sin \varphi ,\cos\theta\right)^T.
\end{eqnarray}
The projections of the effective magnetic noise in Eq.~(\ref{dB1}) on the $\hat{X}$, $\hat{Y}$ and $\hat{Z}$ axis are
\begin{eqnarray}
\delta {B}_{X}&=&b_0\left[\beta_{-}\dot{E}_{cy}\cos\theta\cos\varphi + \beta_{+}\dot{E}_{cx}\cos\theta\sin\varphi\right],\\
\delta {B}_{Y}&=&b_0\left[-\beta_{-}\dot{E}_{cy}\sin\varphi + \beta_{+}\dot{E}_{cx}\cos\varphi\right],\\
\delta {B}_{Z}&=&b_0\left[\beta_{-}\dot{E}_{cy}\sin\theta\cos \varphi + \beta_{+}\dot{E}_{cx}\sin\theta\sin \varphi\right],
\end{eqnarray}
where, $b_0=2e/g\mu _{B} \omega _{d}^{2}$ is defined for simplicity.

One interesting feature here is that under some conditions the cross correlations, such as $S_{YZ}^{+}$ for $\theta=\pi/2$, do not vanish.
However, at the lowest order of $\Gamma/\omega _{Z}$,
as shown in Appendix \ref{append::spindecoher},
the relaxation and dephasing formulae retain the usual form  (typically $\Gamma \ll \omega_Z$, i.e. the decoherence rate is much less than the Zeeman splitting).

Longitudinal fluctuations lead to pure dephasing of the spin qubit, with a dephasing rate of
\begin{equation}
\frac{1}{T_{\varphi }}=\left. S_{ZZ}(\omega)\right|_{\omega\rightarrow0}.
\end{equation}
The noise spectrum usually goes to zero in the limit of zero frequency, so that pure dephasing is often negligible. The $1/f^a$ charge noise, which has significant contribution at low frequencies, could have finite contribution to the dephasing rate (see Appendix \ref{append::puredephas}). Below we will focus on the relaxation effects.

Transverse fluctuations lead to the relaxation of the spin qubit.  The relaxation rate is
\begin{eqnarray}
\frac{1}{T_{1}} &=& 2\left[ \frac{e}{\hbar\omega_d^2}\right] ^{2}F_{SO}(\theta,\varphi)\omega_Z^2S_E(\omega_Z),\label{1T1} \\
F_{SO}&=&(\beta ^{2}+\alpha ^{2}) ( 1+\cos ^{2}\theta)+2\alpha \beta \sin ^{2}\theta \cos 2\phi, \label{F_SO}
\end{eqnarray}
where, $S_{E}(\omega)$ is the Fourier transform of the correlation of the random electric field.  Below we examine the qualitative features of the relaxation rate given here.

The SO interaction dependence of $1/T_{1}$ is contained in $F_{SO}$ in terms of $\alpha $ and $\beta$, the Rashba and Dresselhaus SO interaction constants. These parameters are materials- and device-specific. In Si, $\beta=0$ because of the bulk inversion symmetry, while in GaAs $\beta_{GaAs} \sim 1000$ m/s, in InAs $\beta_{InAs} \sim 30000$ m/s, depending on the structure of the samples.\cite{Studer2010,Sanada2011,Zumbuhl2002,Silva1997} In nanostructures made from either material, $\alpha$ is generally finite. Its magnitude depends on how strongly heterogeneous the underlying quantum well structure is.

The dependence on the direction of the applied magnetic field $\boldsymbol{B}_0$ by $1/T_{1}$ is also contained in $F_{SO}$, in terms of the polar and azimuthal angles $\theta$ and $\phi$. When the polar angle $\theta = 0$, the applied field is along the growth direction of the 2D quantum dot, and $F_{SO}=2(\beta^{2}+\alpha ^{2})$.  It is always larger than that for $\theta=\pi/2$ (in-plane field), when
\begin{equation}
F_{SO}(\theta=\pi/2,\phi)=\beta ^{2}+\alpha ^{2}+2\alpha \beta \cos 2\phi. \label{Fsopara}
\end{equation}
Therefore, if the magnetic field has the same magnitude, the relaxation rates for the in-plane field cases are always slower than the perpendicular case.

The spin relaxation rate $1/T_{1}$ has a sinusoidal dependence on the azimuthal angle $\phi $ of $\boldsymbol{B}_{0}$. Note that Eq.~(\ref{Fsopara}) describes the distance of two vectors $\boldsymbol{a}$ and $\boldsymbol{b}$ with the magnitudes being proportional to $\left\vert \alpha\right\vert $ and $\left\vert \beta \right\vert $ and the angle between them being $\pi -2\phi$ (if $\alpha\beta>0$). The minimum rate is obtained when the two vectors are along the same directions,
\begin{equation}
\left(1/T_{1}\right)_{\mathrm{min}}=2\left[ e\left( |\beta| -|\alpha| \right) /\left(\hbar \omega _{d}^{2}\right) \right] ^{2}\omega_Z^2S_{E}(\omega_Z).
\end{equation}
In the special case when $\alpha =\beta $ and $\phi =\pi /2$ (or $\alpha =-\beta $ and $\phi =0$), $1/T_{1}=0$. In other words, spin relaxation due to charge noise vanishes if $\boldsymbol{B}_{0}$ is along $y$ for $\alpha =\beta $ (or along the $x$ axis for $\alpha =-\beta $). {Such special cases ($\alpha =\pm \beta $) have been discussed previously in the context of spin relaxation due to phonon emission.\cite{Schliemann2003, Golovach2004}} Note that Hamiltonian (\ref{H}) conserves the spin component $\sigma _{y(x)}$ for $\alpha =\beta $ ($\alpha=-\beta $) and $\boldsymbol{B}_{0} \parallel y\,(x)$. This spin conservation results in $T_{1}$ being infinite to all orders in $H_{SO}$.

The spin relaxation rate (\ref{1T1}) has a strong dependence on the QD confinement, $1/T_{1}\propto 1/\omega _{d}^{4}$. Thus this spin relaxation channel can be suppressed by having a strong QD confinement. The dependence on the magnitude of the magnetic field is contained in $\omega_Z^2 S_{E}(\omega_Z)$, which is noise-spectrum-dependent.

\section{Evaluation of Charge Noise induced Spin Relaxation}

Below we present numerical results on the spin relaxation rates for three different noises, namely, $1/f^a$ charge noise, Johnson noise and EWJN. For each electric noise, we carry out numerical calculations on three representative QD structures, namely Si, GaAs, and InAs QD. We consider Si and GaAs QD to be gate-defined QD with the confinement energy $\hbar\omega_d=1$ meV; and InAs QD to be SAQD with smaller size or stronger confinment, i.e. $\hbar\omega_d=4$ meV. In Si, we use the g-factor $g=2$, the electron effective mass $m^{\ast }=0.19m_{0}$, where $m_{0}$ is the free electron rest mass. The Dresselhaus and Rashba SO interaction strength are chosen as $\beta_{Si}=0$ m/s and $\alpha_{Si}=5$ m/s.\cite{Wilamowski2002,Tahan2005,Prada2011} In GaAs, we use $g=-0.44$, $m^{\ast}=0.067m_{0}$, $\beta_{GaAs}=1000$ m/s.\cite{Studer2010,Sanada2011,Zumbuhl2002} In InAs, we use $g=-6.5$, $m^{\ast}=0.023m_{0}$, $\beta_{GaAs}=26900$ m/s.\cite{Knap1996,Silva1997,Nowak2011,Takahashi2010} In both GaAs and InAs QDs, we use $\alpha=0$ m/s for simplicity, although in reality, it could be as large as $500$ m/s in GaAs and $1000$ m/s in InAs.\cite{Zumbuhl2002,Knap1996} As we have discussed above, the relaxation rate when $\alpha$ is finite would depend on the orientation of the applied magnetic field. Except in the highly unlikely case of $\alpha = \beta$, the field-direction-dependence only changes the relaxation rate in the O(1) order.

%We carry out numerical calculations on two representative QD structures, one in GaAs/Al$_{1-x}$Ga$_{x}$As, the other in Si/SiGe. In both cases, the dot confinement energies are set at $\hbar \omega _{d}=1$ meV.  The rest of the parameters are chosen as $a=1$, $l_{0}=100$ nm, and $\sigma _{V}\sim 0.1$ $\mu $eV [so that $A=1$ (V/m)$^{2}$]. For the GaAs QD, we use the bulk g-factor $g=-0.44$, and the electron effective mass $m^{\ast}=0.067m_{0}$, where $m_{0}$ is the free electron rest mass. For the Si QD, we use $g=2$ and $m^{\ast }=0.19m_{0}$.

%In the calculation we choose $\alpha_{Si}=5$ m/s for Si. In GaAs, we choose $\alpha_{GaAs}=0$ m/s for simplicity,

\subsection{$1/f^a$ Charge Noise}

\begin{figure}[tb]
\includegraphics[scale=0.4]{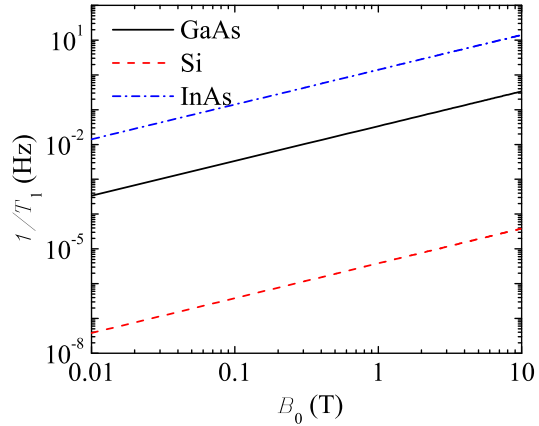}
\caption{Spin relaxation rate $1/T_{1}$ as a function of the magnetic field (in-plane) due to $1/f$ noise in GaAs and Si gate-defined QD ($\hbar\omega_d=1$ meV) and in InAs SAQD ($\hbar\omega_d=4$ meV).}\label{Fig_1f}
\end{figure}

The spin relaxation rate due to the $1/f^a$ charge noise is given by
\begin{equation}
\frac{1}{T_{1}}=2A\omega _{Z}^{2-a}\left[ \frac{e}{\hbar \omega _{d}^{2}}\right] ^{2}F_{SO}(\theta,\phi).\label{1T1_1f}
\end{equation}
The dependence of $1/T_{1}$ on the applied magnetic field is determined by the specific noise spectrum of charge noise, $1/T_{1}\propto B_{0}^{2-a}$. Specifically, if $a=1$ ($S_E \propto 1/\omega$), $1/T_{1}$ depends linearly on the $B_{0}$.

%The rest of the parameters are chosen as $a=1$, $l_{0}=100$ nm, and  For the GaAs QD, we use the bulk g-factor $g=-0.44$, and the electron effective mass $m^{\ast}=0.067m_{0}$, where $m_{0}$ is the free electron rest mass. For the Si QD, we use $g=2$ and $m^{\ast }=0.19m_{0}$.

Fig.~\ref{Fig_1f} shows the spin relaxation rate $1/T_{1}$ due to $1/f$ charge noise as a function of the magnitude of magnetic field for the Si, GaAs and InAs QDs. We choose $a=1$, $l_0=100$ nm, $A=1$ (V/m)$^{2}$ (or $\sigma _{V}\sim 0.1$ $\mu $eV) for $1/f$ noise. The dot confinement energy is set as $\hbar\omega_d=1$ meV for Si and GaAs gate-defined QD and $\hbar\omega_d=4$ meV for InAs SAQD. As shown in the figure, at $B=1$ T, $T_{1}$ is about 10 s for a GaAs QD. For a Si QD $T_{1}\sim 100,000$ s because of the weaker SO interaction. For an InAs SAQD $T_{1}\sim 1$ s due to the combined effect of stronger SO interaction and QD confinement. With $a=1$, the curves here show simple linear dependence on $B_{0}$, a much weaker magnetic field dependence compared with the case of phonon noise, which has $B_0^5$ (or $B_0^7$) dependence for piezoelectric (or deformation) phonon potential.

\subsection{Johnson Noise}

The spin relaxation rate due to Johnson noise from the nearby metallic gate takes the form
\begin{equation}
\frac{1}{T_{1}}=2\left[ \frac{e}{\hbar \omega _{d}^{2}}\right]^{2} F_{SO}(\theta,\phi) \omega _{Z}^{2}S_{V}(\omega_Z)/(el_{0})^{2},\label{1T1_Johnson}
\end{equation}
where $S_{V}(\omega)$ is given by Eq.~(\ref{Sv_John}), and $l_0$ is the length scale between the metal gates that define the QD.

\begin{figure}[tb]
\includegraphics[scale=0.4]{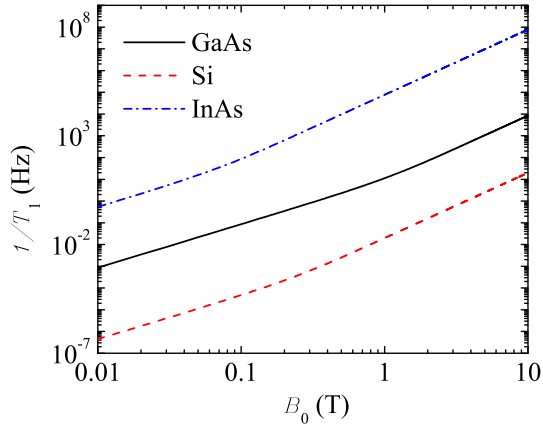}
\caption{Spin relaxation rate $1/T_{1}$ as a function of the magnetic field (in-plane) due to Johnson noise in GaAs and Si gate-defined QD ($\hbar\omega_d=1$ meV) and in InAs SAQD ($\hbar\omega_d=4$ meV).}\label{Fig_Johnson}
\end{figure}

The dependence of $1/T_{1}$ on the applied magnetic field is determined by the factor $\omega _{Z}^{3}\coth(\hbar\omega_Z/2k_{B}T)$, assuming that the cutoff frequency for the Johnson noise is much larger than the Zeeman frequency, $\omega_R \gg \omega_Z$. When the Zeeman energy $\hbar\omega_Z$ is much larger than the thermal energy $k_BT$, we have $\coth(\hbar\omega/2k_BT) \approx 1$, then the spin relaxation rate $1/T_{1}$ has a $B_0^3$ dependence, which has been obtained before theoretically. \cite{Marquardt2005, SanJose2006, SanJose2007} On the other hand, when $\hbar\omega_Z \ll k_B T$ (the white noise limit for the electric fluctuation), we have $\coth(\hbar \omega_Z / 2k_BT) \approx 2k_B T / \hbar \omega_Z$, then the spin relaxation rate $1/T_{1}$ has a $B_0^2$ dependence and linearly proportional to the temperature. The different B-field and temperature dependence for different temperatures here is similar to the phonon induced spin relaxation. \cite{Hanson2007, Zwanenburg2013}

%When the Zeeman frequency is much less than the cutoff frequency, $\omega_Z \ll \omega_R$, the dependence of $1/T_{1}$ on the applied magnetic field is determined by the factor $\omega _{Z}^{3}\coth(\hbar\omega_Z/2k_{B}T)$.  However, when $\omega_Z \gg \omega_R$, the functional dependence changes to $\omega _{Z}\coth(\hbar\omega_Z/2k_{B}T)$ due to the high frequency cutoff from filtering.  In this work we focus on Johnson noise from the outside circuit at room temperature ($T=300$ K). For the magnetic field range we are interested in, the condition $\hbar\omega_Z\ll k_BT$ is always satisfied, so that $\coth(\hbar \omega_Z / 2k_BT) \approx 2k_B T / \hbar \omega_Z$.  Therefore, when $\omega_Z \ll \omega_R$, the relaxation rate $1/T_{1}$ has a $B_0^2$ dependence; when $\omega_Z \gg \omega_R$, $1/T_{1}$ shows no dependence on $B_0$.

Fig.~\ref{Fig_Johnson} gives spin relaxation rate $1/T_{1}$ due to Johnson noise of the metallic gates as a function of the applied magnetic field. We use $T=0.15$ K, $R=50$ $\Omega$, and $\omega_R=10^{15}$ 1/s. As shown in the figure, at $B=1$ T, $T_{1}$ is about $0.1$ s for a GaAs QD;  $T_{1}\sim 100$ s for Si QD, and $T_{1}\sim 10$ $\mu$s for InAs SAQD. The curves show the low-field-to-high-field transition from the $B_{0}^2$ dependence to the $B_{0}^3$ dependence as $B_0$ increases, which is due to the contribution of $\coth(\hbar \omega_Z / 2k_BT)$. The transition occurs at $B_T=k_BT/g\mu_B$, which is $B_T\sim$ 1 T in GaAs, $B_T\sim$ 0.15 T in Si, and $B_T\sim$ 0.05 T in InAs.

%On the other hand, the curve for the Si QD below $B_0 = 1$ T is already in the high field limit (with $\omega_Z \sim 100 \mu$eV while $k_B T \sim 10 \mu$eV), while the bending of the curve above $B_0 = 1$ T is due to the cutoff function, as $\omega_Z$ gradually approaches the value of $\omega_R \sim 300 \mu$eV.

%the curve for the Si QD below $B_0 = 1$ T is already in the high field limit (with $\omega_Z \sim 10^{12}$ 1/s per Tesla). Therefore, in the magnetic field range shown in the figure, $1/T_{1}$ shows no dependence on $B_0$ for Si.

\subsection{Evanescent Waves Johnson Noise}

\begin{figure}[tb]
\includegraphics[scale=0.44]{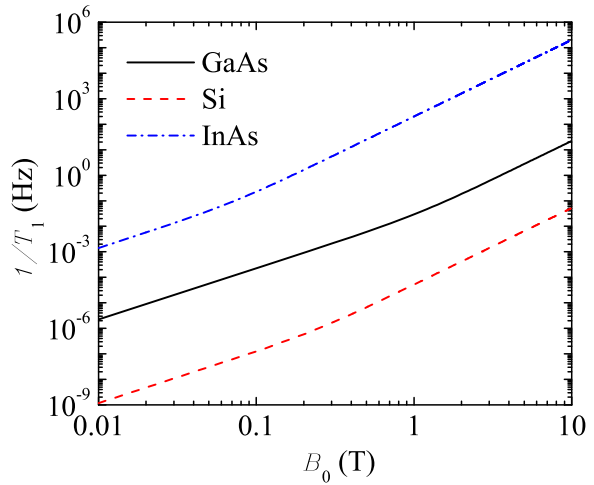}
\caption{Spin relaxation rate $1/T_{1}$ as a function of the magnetic field (in-plane) due to EWJN in GaAs and Si gate-defined QD ($\hbar\omega_d=1$ meV) and in InAs SAQD ($\hbar\omega_d=4$ meV).} \label{EvanWave}
\end{figure}

As we have discussed before, there are two decoherence mechanisms arising from the EWJN, one from the magnetic noise of EWJN, the other is due to the electric noise. The spin relaxation due to EWJN through the direct magnetic dipole interaction has been discussed before.\cite{Langsjoen2012,Poudel2013} Here we estimate the relaxation due to the electric field of EWJN. The relaxation rate is given by,
\begin{equation}
\frac{1}{T_{1}} = \frac{e^2 \omega_Z^3}{8z^3 \sigma \hbar \omega _{d}^{4}}\coth \left( \hbar \omega_Z /2k_{B} T \right) F_{SO}(\theta, \phi), \label{1T1_EWJNE}
%\frac{1}{T_{1}^{\mathrm{SO}}} = \frac{e^2 \omega_Z^3}{8z^3 \sigma \hbar \omega _{d}^{4}}\coth \left( \hbar \omega_Z /2k_{B} T \right) F_{SO}(\theta, \phi), \label{1T1_EWJNE}
\end{equation}
where $\omega_Z$ is the Zeeman frequency, $\sigma$ is the conductance of the metal gates, and $z$ is the distance between the QD and metallic gates.

Eq. (\ref{1T1_EWJNE}) shows that the spin relaxation rate $1/T_{1}$ is inversely proportional to the gate conductivity $\sigma$. It is also inversely proportional to the 3rd power of the distance $z$ between the metallic gates and the QD, $1/T_{1}\propto 1/z^3$, as long as $100$ nm $\leq z\leq$ $\delta/10$ so that the local electrodynamics and quasistatic approximation is valid. In short, this mechanism is not important when the gate is far away from the QD.

The dependence of $1/T_{1}$ on the applied magnetic field is again determined by the factor $\omega _{Z}^{3} \coth( \hbar \omega_Z / 2k_{B} T)$, which is similar to the case of far field Johnson noise. Thus, depending on whether the low field $\hbar \omega_Z \ll k_B T$ or high field $\hbar \omega_Z \gg k_B T$ limit is realized, the relaxation rate $1/T_{1}$ has either $B_0^2$ or $B_0^3$ dependence.

In Fig.~\ref{EvanWave}, the relaxation rates due to the EWJN and SO interaction are plotted as a function of the magnetic field for Si, GaAs and InAs QD. We use $T=0.15$ K for the temperature of the gates, $\sigma=6\times10^7$ S/m for the conductivity of the cooper gate and the distance $z=100$ nm. As shown in the figure, at $B=1$ T, $T_{1}$ is about is about $10$ s for a GaAs QD;  $T_{1}\sim 10^4$ s for Si QD, and $T_{1}\sim 10$ ms for InAs SAQD. The curves show the low-field-to-high-field transition from the $B_{0}^2$ dependence to the $B_{0}^3$ dependence as $B_0$ increases. The transition occurs at the same points as in the Johnson noise case.

%Panel (a) of Figure~\ref{EvanWave} is for GaAs QD, where the crossing from SO interaction mechanism to magnetic dipole interaction mechanism occurred. Panel (b) of Figure~\ref{EvanWave} is for Si QD, where the relaxation due to the magnetic dipole interaction is dominant over the relaxation from the electric EWJN through the SO interaction.

\section{Comparison of Different Noises}

We can now compare the magnetic field dependence of spin relaxation rate for charge noise (including $1/f$ noise, Johnson noise and EWJN) and phonon noise. We will discuss GaAs gate-defined QD, Si gate-defined QD and InAs SAQD, respectively.

\subsection{GaAs and Si gate-defined QDs}

\begin{figure}[]
\includegraphics[scale=0.44]{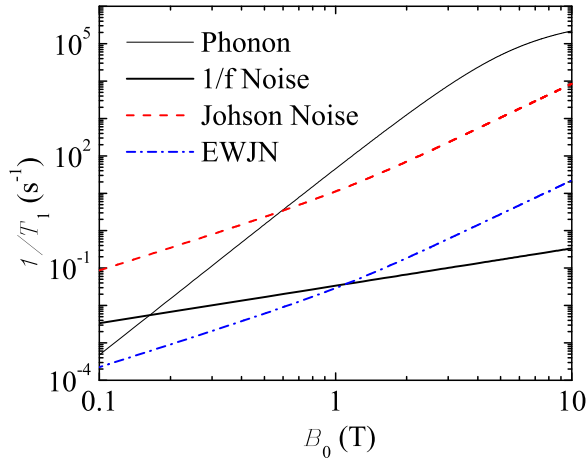}
\caption{Spin relaxation rate $1/T_{1}$ as a function of the applied magnetic field due to different noises in gate-defined defined GaAs QD ($\hbar\omega_d$= 1meV). Data of phonon induced spin relaxation is from Ref. [\onlinecite{Golovach2004}].} \label{Gamma_comp}
\end{figure}

It is well known that the spin relaxation rate in a gate-defined GaAs QD is mainly due to the phonon noise in high magnetic field, where the relaxation rate has a $B_0^5$ (or $B_0^7$) dependence for the piezoelectric (or deformation) phonon potential. \cite{Khaetskii2001, Golovach2004,Amasha2008} In Fig.~\ref{Gamma_comp}, we show the spin relaxation rate as a function of the applied magnetic field due to charge noise and phonon noise in GaAs QD, where the results of phonon noise is from Ref. \onlinecite{Golovach2004}. As shown in the figure, the relaxation rate due to charge noise is less important than the phonon noise in the high B-field regime. However, as the magnetic field decreases, the dominant spin relaxation channel could cross over from phonon noise to charge noise, so that the spin relaxation rate in the low magnetic field deviate from the $B_0^5$ curve (due to piezoelectric phonon), which is consistent with recent experimental observations. \cite{Amasha2008}

%As shown in the figure, the dominant channel could be the Johnson noise through the SO interaction and the EWJN through the magnetic dipole interaction.

\begin{figure}[]
\includegraphics[scale=0.44]{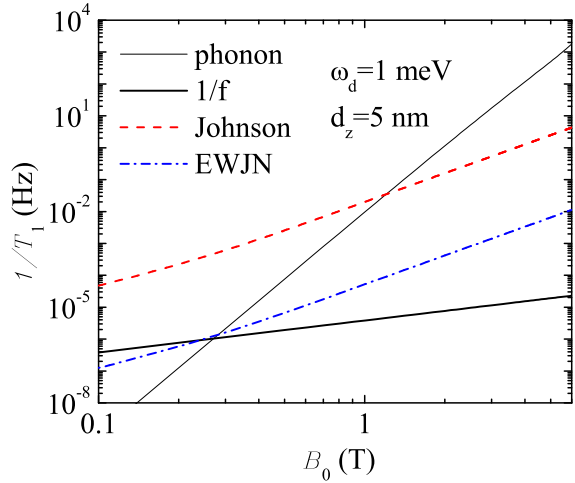}
\caption{Spin relaxation rate $1/T_{1}$ as a function of the applied magnetic field due to phonon noises and charge noise in a gate-defined Si QD with $\hbar\omega_d=1$ meV. The results of phonon induced spin relaxation is calculated based on Ref. [\onlinecite{Golovach2004}].} \label{Gamma_Si}
\end{figure}

In gate-defined Si QDs, spin relaxation has been explored in recent experiments.\cite{Xiao2010,Yang2013} In high magnetic fields, spin relaxation is again dominated by phonon emission.\cite{Xiao2010, Yang2013,Tahan2013}  Compared to GaAs, a particular complexity in Si is the presence of valleys and valley states in quantum wells and QDs.\cite{Hao2013,Gamble2013,Yang2013,Laird2013,Culcer2012}  Here we assume that the valley splitting is large, so that the intra-valley spin-orbit mixing plays the dominant role in spin relaxation.\cite{Yang2013} In Fig.~\ref{Gamma_Si}, we present the spin relaxation rate as a function of the applied magnetic field $B_0$ due to phonon noise and charge noise. The result of phonon noise is calculated based on the formalism from Ref.~[\onlinecite{Golovach2004}].  In Si, relevant parameters include speeds of sound at $s_1\approx 9.33\times 10^5\, \mathrm{cm}/\mathrm{s}$ and $s_2=s_3\approx 5.42\times 10^5\,\mathrm{cm}/\mathrm{s}$, and the mass density of $\rho_c=2.33$ g/cm$^3$.  The dilation and shear deformation potential constants are $\Xi_d = 5$ eV and $\Xi_u=8.77$ eV.\cite{Hu2011,Tahan2013}  Lastly, the device temperature is $T=0.15$ K and the vertical confinement length is $d_z=5$ nm. As shown in Fig.~\ref{Gamma_Si}, relaxation due to phonon emission dominates in the high magnetic field regime, where the rate shows a $B_{0}^7$ dependence, while the charge noise dominates in the low magnetic field regime, similar to the case of GaAs QD.

%, and the results are consistent with recent experimental measurement. \cite{Xiao2010, Yang2013}

\subsection{InAs Self-assembled QD}

InAs SAQDs form on a substrate of GaAs because the lattice mismatch between InAs and GaAs.  As such they are generally much smaller in size compared to the gate-defined dots.  Furthermore, since the formation of InAs SAQDs is sensitive to the growth conditions, their size is difficult to control precisely.  Consequently the QD confinement energy could vary in a wide range, from a few meV to a few tens of meV.  This variation leads to significant modifications to spin relaxation as a result of the strong dependence of spin relaxation rate on the dot confinement energy $\omega_d$.

Another important parameter concerning spins is the g-factor, which is much larger in InAs than in GaAs, with $g=-15$ in bulk InAs crystal and $g=-8$ in an InAs quantum well.\cite{Pidgeon1967, SmithIII1987}  In an InAs SAQD, the g-factor could be very different from that in the bulk due to the dot confinement.\cite{Pryor2006}  The magnitude of the g-factor is dependent on the Ga content and the degree of strain, and ranges widely, from 0.5 to 6.5 in InAs or In$_x$Ga$_{1-x}$As SAQDs.\cite{Kroutvar2004, Heiss2010, Press2010, Takahashi2013, Kanai2011, Takahashi2010} In a gate-defined InAs nanowire QD (NWQD), the measured magnitude of g-factor ranges from 7 to 8.\cite{Schroer2011, Stehlik2013, Stehlik2013} Here we focus on the SAQD, and consider specifically two types of InAs SAQD, the small dots with $g=0.5$ and a confinement energy of $\hbar\omega_d=30$ meV,\cite{Heiss2010, Press2010} and the large dots with $g=-6.5$ and a confinement of $\hbar \omega_d = 4$ meV.\cite{Takahashi2010, Takahashi2013}

%\begin{figure}[]
%\includegraphics[scale=0.44]{phonon_InAs_wd30_dz1nm.eps}
%\caption{Spin relaxation rate $1/T_{1}$ as a function of the applied magnetic field due to phonon noises in InAs self assembled QD with $\hbar\omega_d=30$ meV. The dashed line is for piezoelectric electron-phonon interaction and the dash-doted line is for the deformation electron-phonon interaction.} \label{InAs_phonon}
%\end{figure}

\begin{figure}[]
\includegraphics[scale=0.44]{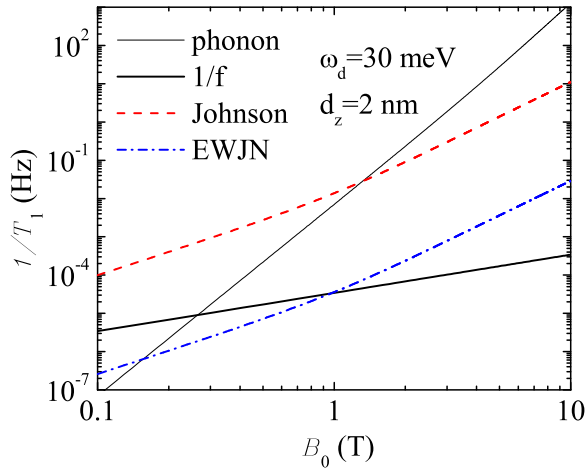}
\caption{Spin relaxation rate $1/T_{1}$ as a function of the applied magnetic field due to different noises in a small InAs SAQD, where the lateral confinement energy is $\hbar\omega_d=30$ meV (equivalent to $d_x=d_y\approx 10$ nm) and the vertical confinement length is $d_z=2$ nm.} \label{InAs_small}
\end{figure}

%In Fig.~\ref{InAs_phonon}, we show the spin relaxation rate as a function of the applied magnetic field in an InAs self-assemble QD due to the phonon noise with the confinement energy of $\hbar\omega_d=30$ meV.
%Here we first evaluate the effects of these two factors on phonon-induced spin relaxation in InAs QDs.

%Furthermore, $\overline{\Xi}_j=\delta_{j,1}\Xi_0$ with $\Xi_0\approx 6\,{eV}$, and $\overline{\beta}_{1,\vartheta}= 3\sqrt{2}\pi h_{14}\kappa^{-1}\sin^2\vartheta\cos\vartheta$, $\overline{\beta}_{2,\vartheta}= \sqrt{2}\pi h_{14}\kappa^{-1}\sin2\vartheta$, $\overline{\beta}_{3,\vartheta}= \sqrt{2}\pi h_{14}\kappa^{-1}(3\cos^2\vartheta-1)\sin\vartheta$, with $h_{14}\approx 0.046\,\mathrm{C}/\mathrm{m}^2$ and $\kappa\approx 15.15$. The piezoelectric phonon dominate the spin relaxation rate at low magnetic field, while the deformation phonon dominates at high magnetic field.

%For GaAs, we use $s_1\approx 4.7\times 10^5\, \mathrm{cm}/\mathrm{s}$ and $s_2=s_3\approx 3.37\times 10^5\,\mathrm{cm}/\mathrm{s}$. Furthermore, $\overline{\Xi}_j=\delta_{j,1}\Xi_0$ with $\Xi_0\approx 7\,{eV}$, and $\overline{\beta}_{1,\vartheta}= 3\sqrt{2}\pi h_{14}\kappa^{-1}\sin^2\vartheta\cos\vartheta$, $\overline{\beta}
%_{2,\vartheta}= \sqrt{2}\pi h_{14}\kappa^{-1}\sin2\vartheta$, $\overline{\beta}_{3,\vartheta}= \sqrt{2}\pi h_{14}\kappa^{-1}(3\cos^2\vartheta-1)\sin\vartheta$, with $h_{14}\approx 0.16\,\mathrm{C}/\mathrm{m}^2$ and $\kappa\approx 13$.

Figure~\ref{InAs_small} shows the spin relaxation rate as a function of magnetic field in a small InAs SAQD due to charge noise and phonon noise. The lateral confinement energy here is $\hbar\omega_d=30$ meV (equivalent to $d_x=d_y\approx 10$ nm), and the vertical confinement length is $d_z=2$ nm. The electron phonon interaction in an InAs SAQD is of the same form as in GaAs due to the similar lattice structure, and the calculation for phonon noise is based on the same procedure as in Ref. \onlinecite{Golovach2004}. For InAs, we use the phonon velocity $s_1\approx 4.28\times 10^5\, \mathrm{cm}/\mathrm{s}$ for longitudinal acoustic modes and $s_2=s_3\approx 2.65\times 10^5\,\mathrm{cm}/\mathrm{s}$ for transverse acoustic modes. Other parameters include the deformation potential $\Xi_0\approx 6\,{eV}$, the piezoelectric constant $h_{14}\approx 0.046\,\mathrm{C}/\mathrm{m}^2$, the relative dielectric constant $\kappa \approx 15.2$, and the density $\rho_c=5.67$ g/cm$^3$.  As shown in Fig.~\ref{InAs_small}, the spin relaxation rate due to phonon noise dominate in the high magnetic field regime. While the charge noise dominate in the low magnetic field regime. \cite{Kroutvar2004, Heiss2010, Press2010}  These small InAs SAQDs are most often probed optically, and normally metallic gates are far away from the QDs in those experiments, so that effects of Johnson noise from the gates are reduced.\cite{Kroutvar2004, Heiss2010, Press2010} In these cases, the spin relaxation rate in low magnetic fields could be dominated by the 1/f noise.

%As shown in the figure, the piezoelectric phonon dominate the spin relaxation rate at low magnetic field, while the deformation phonon dominates at high magnetic field. In the high magnetic field, the spin relaxation also shows plateau as in GaAs QD.\cite{Golovach2004}

\begin{figure}[]
\includegraphics[scale=0.44]{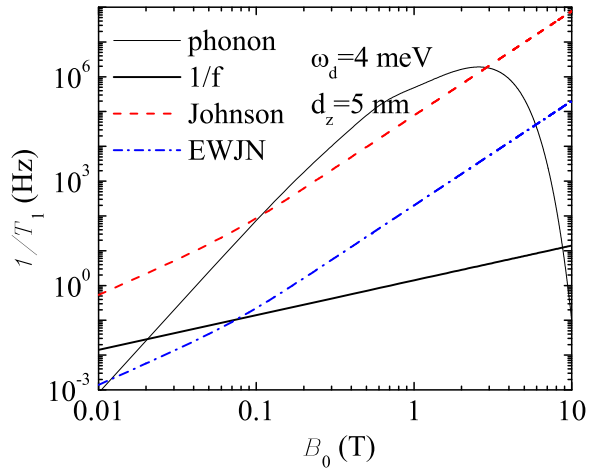}
\caption{Spin relaxation rate $1/T_{1}$ as a function of the applied magnetic field due to different noises in a large InAs SAQD, where the lateral confinement energy is $\hbar\omega_d=4$ meV (equivalent to $d_x=d_y\approx 30$ nm) and the vertical confinement length is $d_z=5$ nm.} \label{InAs_large}
\end{figure}

In Fig.~\ref{InAs_large} we plot the spin relaxation rate as a function of the applied magnetic field in a large InAs SAQD, which has a lateral confinement energy of 4 meV (corresponding to a confinement length of ~30 nm) and a vertical confinement length of 5 nm. \cite{Takahashi2010} Equation~(\ref{1T1}) shows that spin relaxation is generally much faster in larger dots due to the smaller confinement energy, which is clearly illustrated by the different vertical scales in Fig.~\ref{InAs_large} as compared to Fig.~\ref{InAs_small}.  Indeed, in the sample considered in Fig. \ref{InAs_large}, spin relaxation time due to Johnson noise could be as short as 10 $\mu$s in a 1 Tesla field.  More interestingly, spin relaxation due to phonon noise is strongly suppressed in high magnetic fields, since a large Zeeman splitting and a large electron wave function mean that the matrix element for the electron-phonon interaction gets averaged out in a large dot, and consequently spin relaxation is suppressed.  In a GaAs gate-defined dot, the phonon induced spin relaxation shows a plateau in the reasonably high fields due to this suppression, while in a large InAs dot, the spin relaxation rate is suppressed more strongly due to the very large g-factor and the relatively small cutoff wave vector.  This strong suppression of phonon induced spin relaxation has been previously observed experimentally for the singlet-triplet relaxation in a two-electron QD,\cite{Meunier2007} and been discussed for single-electron spin in a InAs NWQD.\cite{Trif2008} As a result of the strong suppression of phonon noise in the high magnetic field regime, the charge noise induced spin relaxation could be dominant at both low and high magnetic fields in a large InAs SAQD, which should be experimentally observable.

\section{discussion}

In the current study we use phonon-induced spin relaxation as the benchmark for comparison when we calculate spin relaxation due to charge noises.  At low magnetic fields, spin relaxation due to two-phonon processes may become important, especially at high temperatures.\cite{Khaetskii2001, Stavrou2006, Trif2009, Fras2012} The corresponding spin relaxation rate could be faster than that from the one-phonon processes due to the vanishingly small phonon density of states at low frequencies.  As such an important question is whether two-phonon processes may be more important than the charge-noise induced relaxation processes, so that the charge noise effects we study in this manuscript would be masked at low fields.  Here we would like to point out that at low temperatures, two-phonon processes are strongly suppressed.  The basic physics here is quite straightforward: at low temperatures, the higher-energy phonons that make two-phonon processes important at higher temperatures are not excited, so that they cannot contribute to the spin relaxation process.

Consider the example of an electron spin in a GaAs QD.  Here the spin relaxation rate due to two-phonon processes through the piezoelectric electron-phonon interaction has a different temperature dependence for temperatures smaller and larger than $T_0=k_B^{-1}\sqrt{mv_j^2\hbar\omega_d}$.\cite{Khaetskii2001}  In a GaAs dot with $\hbar\omega_d=1$ meV, $T_0\approx1$ K. At low temperatures when $T<T_0$, the spin relaxation rate is given by\cite{Khaetskii2001}
\beq
\Gamma^{(2p)}(B)=\frac{\Lambda_p^2}{\hbar}\sum_j\frac{s_j^2}{\beta^2}\frac{(g\mu_BB)^2(m^*s_j^2)^{5/2}}{(\hbar\omega_d)^{7/2}} \left(\frac{T}{T_0}\right)^{9},
\eeq
where $s_j$ is the phonon velocity of branch-$j$, and $\Lambda_p$ is the effective spin-piezoelectric phonon coupling strength. In a GaAs QD, $\Lambda_p\sim 10^{-2}$. \cite{Khaetskii2001} For a working temperature of $T \sim 0.15$ K in spin qubit experiments, the two-phonon rate $\Gamma^{(2p)}(B)=10^{-6}B^2$ s$^{-1}$. Therefore, the spin relaxation due to two-phonon processes is negligible at low temperatures compared with the other mechanisms we have discussed so far. In other words, based on our calculations, at low fields and low temperatures, spin-orbit interaction and charge noise (1/f and Johnson noises) should provide the dominant spin relaxation channel, as illustrated previously.

On the other hand, in the case of InAs SAQDs, where many existing experiments are done at higher temperatures, spin relaxation at low magnetic field could be dominated by two-phonon processes in those experiments.  However, at lower temperatures that a spin qubit is operated, our results should still hold for the InAs dots.

% or the Johnson noise from the big Schottky gate that controls the number of electrons in the dot.

Finally, we would like to emphasize that the relative strength of each relaxation mechanism always varies with different parameters. First, the relaxation rate due to the mechanisms through the SO interaction could vary due to the different values of Dresselhaus SO constant $\beta$, and could be $(1+|\alpha/\beta|)^2$ times larger, or $(1-|\alpha/\beta|)^2$ times less, depending on the Rashba SO constant $\alpha$ and the orientation of the applied magnetic field. Furthermore, Johnson noise could be even more important when the resistance of the circuits becomes larger and $1/f$ noise could also be important with larger noise magnitude.

\section{conclusion}

In conclusion, we have studied spin decoherence of a quantum-dot-confined electron due to charge noise. We focus on the spin decoherence originates from the SO interaction and momentum scattering due to charge noise.  We find that both relaxation and pure dephasing are present in our calculation, although the latter is very weak in general for charge noise. We find that, in a gate-defined GaAs or Si QD, the dominant spin relaxation channel could crossover from phonon noise to charge noise, e.g. Johnson noise, as the magnetic field decreases below 1 Tesla. In a small InAs self-assembled dot, 1/f noise could be the dominant spin relaxation source at low magnetic field if no metallic gate is attached to the dot.  In a large InAs dot, spin relaxation due to phonon noise should be strongly suppressed in high magnetic fields, so that spin relaxation due to charge noise could dominate in both low and high magnetic fields. Quantitatively, in a 1 Tesla field, the spin relaxation time due to typical charge noise is about $100$ s in Si, $0.1$ s in GaAs for a gate-defined QD with a $1$ meV confinement, and $10$ $\mu$s in InAs SAQD with a $4$ meV confinement.

%In other words, this is a relaxation dominated decoherence channel, with $T_{2}=2T_{1}$ at the lowest order, similar to the phonon noise case.
%The relaxation rate is inversely proportional to the fourth power of the confinement energy, so that spin decoherence is faster for larger quantum dots.

We thank support by US ARO (W911NF0910393) and NSF PIF (PHY-1104672).

\appendix
\section{Properties of the Superoperator $L_d$}
\label{append::Ld}

In this Appendix we describe properties of the superoperator $\mathbb{L}_d$, and sketch how we obtain the result of Eq.~(\ref{LdHso}) and (\ref{xi}).
Recall that the superoperators $\mathbb{L}_{d}$ is defined as $\mathbb{L}_{d}A\equiv \lbrack H_{d},A]$, $\forall A$.  The properties of $\mathbb{L}$ are different from that in the previous works,\cite{Golovach2004,Borhani2006,Huang2013} since our dot Hamiltonian $H_d$ contains the charge noise in the form of a time-dependent dot position,
\begin{equation}
H_{d}=\frac{{\pi }^{2}}{2m^{\ast }}+\frac{1}{2}m^{\ast }\omega_{d}^{2}\left[ \boldsymbol{r}-\boldsymbol{R}\left( t\right) \right]^{2}.
\end{equation}
Since we are interested in obtaining $\mathbb{L}_d^{-1} H_{SO}$, and $H_d$ commutes with spin operators, the following relations involving $\mathbb{L}_{d}$ are the only ones relevant for our calculation:
\begin{eqnarray}
&&-\mathbb{L}_{d}\left[x-R_{x}\left( t\right) \right] = \left[ x,{\pi _{x}^{2}}/{2m^{\ast }}\right] =i\hbar{\pi _{x}}/{m^{\ast }}, \\
&&-\mathbb{L}_{d}\left[y-R_{y}\left( t\right) \right] = \left[ y,{\pi _{y}^{2}}/{2m^{\ast }}\right] =i\hbar{\pi _{y}}/{m^{\ast }}, \\
&&-\mathbb{L}_{d}\pi _{x} = -i\hbar m^{\ast }\omega _{d}^{2}\left[x-R_{x}\left( t\right) \right] -i\hbar \omega _{c}\pi _{y}, \\
&&-\mathbb{L}_{d}\pi _{y} = -i\hbar m^{\ast }\omega _{d}^{2}\left[y-R_{y}\left( t\right) \right] +i\hbar \omega _{c}\pi _{x},
\end{eqnarray}
where, $\omega _{c}\equiv e B_{0z} / m^{\ast } c$ is the cyclotron frequency of the electron in the 2DEG in the presence of the magnetic field whose perpendicular magnitude is $B_{0z}$. The above equations can be written in the matrix form: $\mathbb{L}_{d}X = M X$, where
\begin{equation}
X\equiv[\pi _{x},\pi _{y},x-R_{x}(t),y-R_{y}(t)]^T. \nonumber
\end{equation}
The inverse of $\mathbb{L}_{d}$ can now be expressed as $\mathbb{L}_{d}^{-1}X=M^{-1}X$, where $M^{-1}$ can be obtained by doing the matrix inversion of $M$,
\begin{equation}
\mathbb{L}_{d}^{-1}X =\frac{1}{i\hbar }\left[
\begin{array}{cccc}
0 & 0 & -m^{\ast } & 0 \\
0 & 0 & 0 & -m^{\ast } \\
\frac{1}{m^{\ast }\omega _{d}^{2}} & 0 & 0 & \frac{\omega _{c}}{\omega _{d}^{2}} \\
0 & \frac{1}{m^{\ast }\omega _{d}^{2}} & -\frac{\omega _{c}}{\omega _{d}^{2}}& 0
\end{array}
\right] X.
\end{equation}%
Using the expressions here for $\mathbb{L}_{d}^{-1}$, it is straightforward to obtain
$
\mathbb{L}_{d}^{-1}H_{SO}=i\left( \boldsymbol{\sigma }\cdot \boldsymbol{\xi }\right) ,
$
where $\boldsymbol{\xi }$ is given by Eq.~(\ref{xi}) in the main text.  It is time-dependent instead of constant, in contrast with that in the previous works.\cite{Golovach2004,Borhani2006,Huang2013}

\section{Spin relaxation and dephasing rate}
\label{append::spindecoher}

%\begin{eqnarray}
%b & = & \sum_i\Gamma_{ii}, \\
%c & = &\omega _{Z}^{2}+\omega _{Z}\left( \Gamma _{XY}-\Gamma _{YX}\right)\notag\\
%  & & + \frac{1}{2}\sum_{ij}(1-\delta_{ij})(\Gamma _{ii}\Gamma _{jj}-\Gamma _{ij}\Gamma _{ji}), \\
%d & = &\omega _{Z}^{2}\Gamma _{ZZ}+\omega _{Z}\sum_{ij}\epsilon_{ijZ}\left( \Gamma _{ij}\Gamma_{ZZ}-\Gamma _{iZ}\Gamma _{Zj}\right)\notag\\
%  & & +\sum_{ijk}\epsilon_{ijk}\Gamma _{Xi}\Gamma _{Yj}\Gamma _{Zk}.
%\end{eqnarray}

In this Appendix we derive the spin relaxation and dephasing rates for a general Bloch Equation, where all the elements are present in the decoherence matrix $\boldsymbol{\Gamma}$.\cite{Borhani2006} We will show that, even though the noise in different directions are correlated, the relaxation and dephasing rate are generally still determined by the noise autocorrelations along the major axes.

The spin relaxation and dephasing rates are determined by the real parts of the solutions of the secular equation for the decoherence matrix det$\left\Vert-\Gamma _{ij}-\lambda \delta _{ij}+\varepsilon _{ijk}\omega _{k}\right\Vert=0$, where $\lambda$ is the eigenvalue of the matrix. The secular equation is a cubic equation
\begin{equation}
\lambda ^{3}+b\lambda ^{2}+c\lambda +d=0.
\end{equation}
If we choose the $Z$-axis to be along the direction of the magnetic field ($\boldsymbol{\omega }\equiv \omega _{Z}\left[0,0,1\right]$), and suppose all the matrix elements $\Gamma _{ij}$ are much smaller than the Zeeman frequency, $\Gamma _{ij}\ll \omega _{Z}$, we have $b=\sum_i\Gamma_{ii}$, $c\approx \omega_{Z}^{2}$ and $d \approx \omega _{Z}^{2}\Gamma _{ZZ}$. The eigenvalues can then be obtained as
$\lambda _{1} \approx -\Gamma _{ZZ}$,
$\lambda _{2}=\lambda_{3}^* \approx -\frac{\Gamma _{XX}+\Gamma _{YY}}{2}+i\omega _{Z}. $
Since each eigenvalue $\lambda_i$ determines the dynamics of each eigenstate, in which the real part corresponds to the decay rate and the imaginary part corresponds to the oscillation frequency. Thus, we identify that the eigenvalue $\lambda_1$ corresponds to relaxation and $\lambda_{2,3}$ correspond to the dephasing process, so that the relaxation and the dephasing rates are determined as
\begin{eqnarray}
1/{T_{1}}&\approx &\Gamma _{{ZZ}},\\
1/{T_{2}}&\approx &({\Gamma _{{XX}}+\Gamma }_{YY})/{2}.
\end{eqnarray}
By substituting the spin decoherence matrix elements $\Gamma_{\mu\nu}$ obtained from the Born-Markov master equation,\cite{Borhani2006} we arrive at
\begin{eqnarray}
\frac{1}{T_{1}} &\approx & S_{XX}^{+}(\omega _{Z})+S_{YY}^{+}(\omega _{Z}) - \sum_{ij} \epsilon_{ijZ}S_{ij}^{-}(\omega _{Z}),\\
\frac{1}{T_{2}} &\approx &\frac{1}{2T_1} + S_{ZZ}^{+}(0),
\end{eqnarray}%
where, the cross correlations $-\sum_{ij} \epsilon_{ijZ}S_{ij}^{-}(\omega _{Z})=S_{YX}^{-}(\omega _{Z})-S_{XY}^{-}(\omega _{Z})$ are from the diagonal terms of spin decoherence matrix, and it can be rewritten as
\beqa
\lefteqn{S_{YX}^{-}(\omega)-S_{XY}^{-}(\omega)}\notag\\
%&&=\frac{g^{2}\mu _{B}^{2}}{\hbar ^{2}}{Re}\int_{0}^{\infty }dt\left\langle \delta B_{Y}(t)\delta B_{X}(0)\right\rangle \sin \left( \omega t\right) dt,\notag\\
%&&-\frac{g^{2}\mu _{B}^{2}}{\hbar ^{2}}{Re}\int_{0}^{\infty }dt\left\langle \delta B_{X}(t)\delta B_{Y}(0)\right\rangle \sin \left( \omega t\right) dt,\notag\\
&&=\frac{g^{2}\mu _{B}^{2}}{2\hbar ^{2}}\int_{-\infty}^{\infty }dt\left\langle [\delta B_{X}(0),\delta B_{Y}(t)]_+\right\rangle \sin \left( \omega t\right) dt.
\eeqa
Therefore, this cross-correlation only has contribution when the function $\left\langle [\delta B_{X}(0),\delta B_{Y}(t)]_+\right\rangle$ is an odd function of time.

In the case of $\left\langle \delta B_{i}(t)\delta B_{j}(0)\right\rangle=\left\langle \delta B_{j}(t)\delta B_{i}(0)\right\rangle$, the expression for spin relaxation and dephasing rates are further simplified
\begin{eqnarray}
\frac{1}{T_{1}} &\approx &S_{XX}^{+}(\omega _{Z})+S_{YY}^{+}(\omega _{Z}), \\
\frac{1}{T_{2}} &\approx &\frac{1}{2T_{1}}+S_{ZZ}^{+}(0),
%\frac{1}{T_{2}} &=&\frac{1}{2T_{1}}+\left.S_{ZZ}^{+}(\omega)\right|_{\omega\rightarrow0},
\end{eqnarray}%
where, $S_{ii}^{+}(\omega)$ can be written as
\begin{equation}
S_{ii}^{+}(\omega )\equiv \frac{g^{2}\mu _{B}^{2}}{2\hbar ^{2}}\int_{-\infty}^{\infty }dt\left\langle \delta B_{i}(t)\delta B_{i}(0)\right\rangle \cos \left( \omega t\right) dt. \nonumber
\end{equation}
Therefore, at the lowest order approximation of $\Gamma_{ij}/\omega_Z$, even though the noise in different directions are correlated, the relaxation and dephasing rate are still determined by the noise autocorrelations $S_{ii}(\omega)$ along the major axes. These results can also help simplify calculations. For example, we can calculate the spin relaxation rate without doing the additional rotations to eliminate the cross-correlations, which was adopted before, for example in Ref. \onlinecite{Golovach2004}.

\section{Pure dephasing rate for $1/f^a$ noise}
\label{append::puredephas}

In this Appendix we evaluate pure dephasing due to $1/f^a$ noise.  The dephasing rate is given by
\begin{equation}
\frac{1}{T_{\varphi }}=2A\left[ \frac{e}{\hbar \omega _{d}^{2}}\right] ^{2}F_{SO}^Z\left. \left[ \omega^{2-a}\right] \right\vert_{\omega \rightarrow 0},
\end{equation}
where, $F_{SO}^Z(\theta,\varphi)=\sin^2\theta\left(\beta^2 + \alpha^2 - 2\beta\alpha\cos\varphi\right).$  In general, we have $a<2$, and $\left. \left[ \omega^{2-a}\right] \right\vert _{\omega\rightarrow 0}$ goes to zero in the limit of zero frequency. We thus expect pure dephasing to be negligible in these cases.  Although pure dephasing could be finite (as compared to relaxation) when $a \geq 2$, the rate is limited due to small noise amplitude.

Quantitatively, the off-diagonal density matrix element for the spin decays in the form $\exp \left( -\varphi \left( t\right) \right)$, where $\varphi \left( t\right) = \int_{0}^{\omega _{c}}d\omega S_{ZZ}\left( \omega\right) \left[{2\sin (\omega t/2)}/{\omega}\right] ^{2}$,\cite{Duan1998,Astafiev2004} or
\begin{equation}
\varphi \left( t\right) =4A\frac{e^2F_{SO}^Z}{\hbar^2\omega _{d}^{4}}\int_{\omega_{0}}^{\omega _{c}}d\omega \frac{\sin^2(\omega t)}{\omega^{a}}, \label{varphi}
\end{equation}
where $\omega _{c}$ is the upper cutoff frequency chosen as $10^9$ 1/s and $\omega_0$ is the lower cutoff frequency chosen as the inverse of the experiment time, nominally at 1 s.  By numerically evaluating Eq.~(\ref{varphi}), we find that the error $1-\exp\left[-\varphi\left( t\right)\right] $ is indeed extremely small for $1/f$ noise ($a=1$) in GaAs, saturating around $10^{-11}$ at the long time limit $t>10^9$ s. Therefore, as expected, the dephasing rate $1/T_{\varphi }$ (defined by $\varphi\left( T_{\varphi }\right) =1$) is negligible compared with the longitudinal relaxation rate $1/T_{1}$.  We thus focus on spin relaxation in this manuscript.

%\bibliographystyle{apsrev}
%\bibliography{chargenoise}

\end{document}